# A Raman spectroscopic study of zircons on micro-scale and Its significance in explaining the origin of zircons

Xuezhao Bao[a1], Huiming Li[b], Songnian Lu[b]

[a]Department of Earth Sciences, the University of Western Ontario, 34-534, Platt's Lane, London, Canada, N6G 3A8

[b]Tianjin Institute of Geology and Mineral Resources, Tianjin, China 300170

**Abstract** The magmatic/igneous and metamorphic zircons were investigated with Raman spectrum microprobe analysis. We found notable differences between these two kinds of zircons exhibited by the variation trend of Raman peak intensity from core to rim of a crystal. In magmatic zircons, the intensity and the ratio ($\Delta$ = H/W) of Raman spectrum peaks (H = intensity; W = half-height width) gradually decrease from core to rim of a crystal, which is produced by an increase in metamictization degree and suggests an increase in U and Th concentrations from core to rim. In metamorphic zircons, there are two kinds of crystals according to their Raman spectra: the first group of zircons exhibits a variation trend opposite to those of magmatic zircons, tending to increase in the Raman peak intensity and $\Delta$ value from core to rim of a crystal, which is produced by a decrease in metamictization degree and indicates a decrease of U and Th concentrations from core to rim of a crystal. The second group of zircons exhibits no change in Raman peak intensity and $\Delta$ value through a crystal. The data of infrared and Raman spectra of these crystals show that they are well crystallized and have no lattice destruction induced by metamictization, and are thought to crystallize in high temperature stages of metamorphism. During these stages, the U and Th ions have been removed by metamorphic fluids from the parent rocks of these zircons. The other difference between magmatic and metamorphic zircons is the background level of their Raman spectra, which is high and sloped in magmatic zircons, but low and horizontal in metamorphic zircons. For the former, this may be caused by impurities with fluorescent radiation due to a higher crystallization temperature than the latter's. The differences between magmatic and metamorphic zircons can be used to identify the genesis of zircons and understand the origin and evolution history of their parent rocks.

## 1. INTRODUCTION

Raman spectrum microprobe is a new technique used to analyze the micro-structure of materials. We have analyzed the micro-structure of zircon crystals, and found that the peak intensity variation trends of Raman spectra from core to rim are different between magmatic and metamorphic zircons (Bao and Gan, 1996), but we have not understood the mechanism producing these differences. Nasdala et al (1996) found that a weakening in peak intensity of Raman spectra of zircons is produced by metamictization-the damage of crystal lattice caused by radiation from U and Th in the crystals. I have found that the U and Th zonation of zircons can be used to identify magmatic and metamorphic zircons (Bao, 1995). Therefore, in this study, we will include more Raman spectrum microprobe data of zircons to establish a new method to identify magmatic and metamorphic zircons.

---
[1] Corresponding author. E-mail address: xuezhaobao@hotmail.com (Xuezhao Bao)

## 2. THE GEOLOGICAL OCCURENCES OF SAMPLES AND LABORATORY WORKS

Among these magmatic zircons, two groups, one from each area, were extracted from plagioclase-amphibolites: metamorphic mafic volcanic rocks. Their magmatic origin is based on their geological settings, zircon morphology, and single zircon U-Th-Pb age (Bao, 1995; Bao and Gan, 1996). Another magmatic zircon sample was separated from a pegmatite sample taken from Xiao Qinlin, China. Of these metamorphic zircons, one of them was a metamorphic re-crystallized zircon from a metamorphic granite sample located in Danzhu, China (Bao, 1995). The other two samples were chosen from the plagio-amphibolite in Mayuan groups, Mayuan, North Fujiang. The parent rock of the two zircon samples is a mafic volcanic rock formed in the early Proterozoic era with a 2.3 Ga ± of U-Th-Pb zircon age. However, the two metamorphic zircons have an age of 0.4

**Table 1 Geological occurrence, U-Th-Pb age, and chemical composition zonation features of zircons**

| genesis | Sample # | location | Rock type | U-Th-Pb age/Ma | morphology | Composition zonation from core to rim of a crystal |
|---|---|---|---|---|---|---|
| metamorph-ic | 87013 | Danzhu, Longquan, zhejiang | Medium-grade metamorphic granodiorite | 1978 ± | Rounded shape and core | Decreasing Hf and U+Th, increasing Zr/Hf ratio[1] |
| | M-b1 | Mayuan Group, Jiang Yan, Fujian province | Medium-grade metamorphic Plagioclase-Amphibolite | 400 ± | Rounded shape and core | Decreasing Hf, increasing Zr/Hf ratio[2] |
| | M-b2 | | | 400 ± | | |
| magmatic | 9303 | Xiao Qinling | Pegmatite | 1850± | long columnar euhedral shape | increasing Hf and U+Th, decreasing Zr/Hf ratio[1] |
| | T9305 | Langfang, hebei | Plagioclase-Amphibolite | | Euhedral magmatic growth zonings | increasing Hf, decreasing Zr/Hf ratio[1] |
| | M-y1 | Same as M-b1 M-b2 | Same as M-b1 M-b2 | 2300 ± | Euhedral magmatic growth zonings | increasing Hf, decreasing Zr/Hf ratio[2] |
| | M-y2 | | | 2300 ± | | |

1. Bao, 1995
2. Bao and Gan, 1996

Ga±, and are a product of a metamorphic event that occurred at 0.4 Ga±. The geological occurrences, composition zonation, and genesis analyses of these zircons were described in previous studies (Bao, 1995; Bao and Gan, 1996).

These samples were mounted in epoxy resin, and then grinded and polished for analyses. A T6400 Laser Ablation Raman spectrometer from France's J.Y. Company was used to analyze these zircons. Analysis conditions: micro-Raman x10 camera lens; CCD reading spectra; accumulated analysis time: 150 seconds/analysis point; optic slit: 200/250; wavelength of laser beam: 514.5 nm; power of laser beam: 200 nw; spot size in analyzed zircons is 2 μm in radius and 2 μm in depth. Infrared spectra were measured with a Specord 75 IR Infrared spectrometer in the lab of Tianjin Institute of Geology and Mineral Resource. The analyzed results are listed in Table 2.

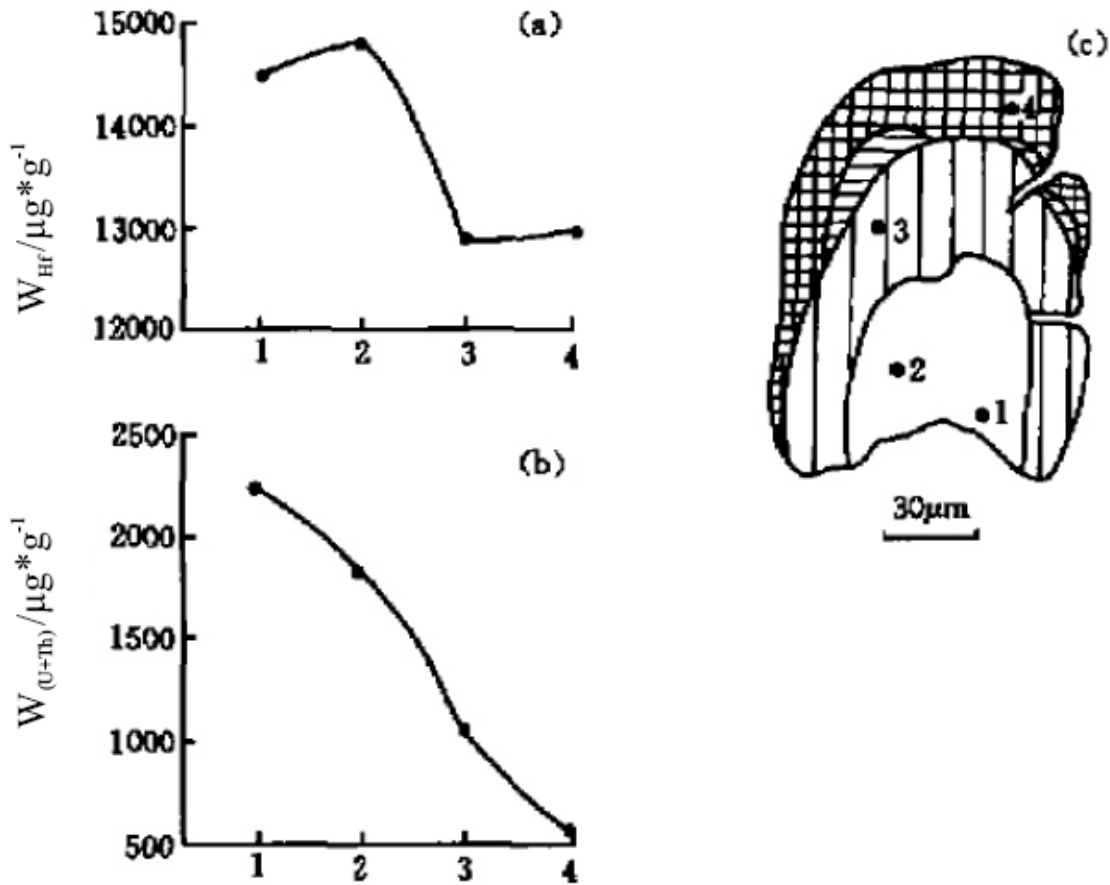

**Fig. 1. The Hf and (U + Th) concentration variations from core to rim (1 →4) of zircon 185/38 and its growth zoning features (based on the data from Navata, 1996)**

## 3. RESULTS AND DISCUSSION

3.1 Genesis determination of zircons

Based on the U-Pb ages, composition zonations, growth zonings and morphologic features combined with their geological settings, we have concluded that zircons T9305 and 9303 (Bao, 1995), and zircons M-y1 and M-y2 (Bao and Gan, 1996) are of magmatic origin, but zircons M-b1 and M-b2 (Bao and Gan, 1996), and 87013 (Bao, 1995) are of metamorphic origin. Zircon 185/38 was chosen from a lightly colored granulite. The data including U, Th and Hf concentrations and the interior morphological picture were extracted from Nasdala et al. (1996) as shown in fig. 1. It clearly indicates that this zircon has a rounded core, and rounded growth zonings, which are consistent with metamorphic zircons (Bao, 1995; Bao et al., 1996). From core to rim of this zircon, Hf and U + Th concentration decreases as indicated in a) and b) in fig.1, which is also consistent with metamorphic zircons (Bao, 1995). In addition, this zircon was chose from a high grade metamorphic rock that has undergone a granulite facies metamorphism; therefore, it is a metamorphic zircon.

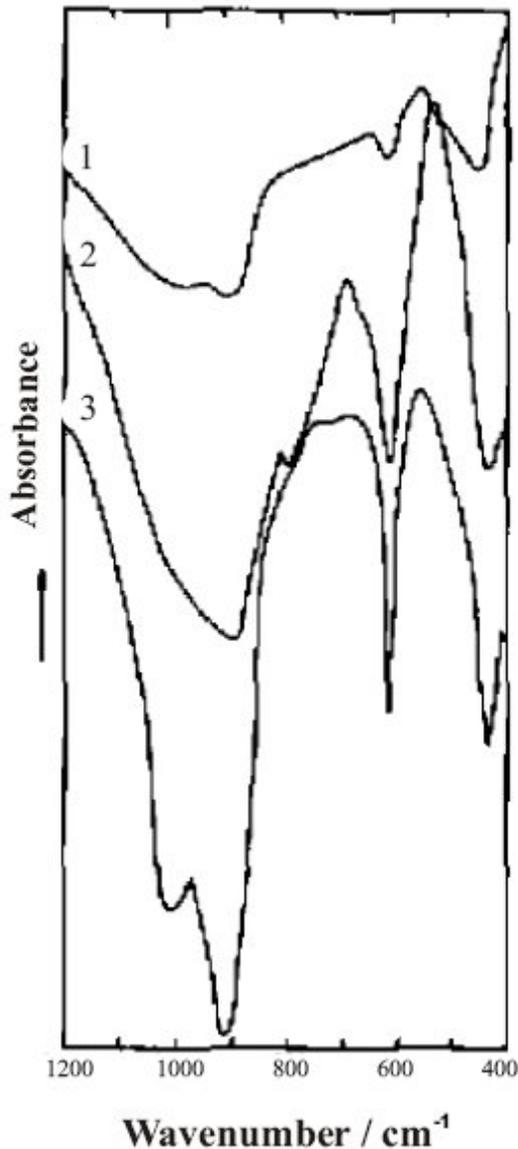

Fig.2 The infrared absorption spectra of multi-crystal zircon samples 9303, T9303 and 87013
1-9303; 2-T9305; 3-87013

3.2 Infrared and Raman spectrum analyses

Gao and Pang (1986) found that with an increase in metamictization degree, the intensity of infrared absorption spectrum peaks decreases, but the width or half-height width of the peaks increases. The metamictization degree can be quantitatively calculated using the values of intensity (H) and half-height width (W) of peak 610 $cm^{-1}$: $\Delta 610 = H/W$. Namely, the smaller the $\Delta 610$, the higher the metamictization degree. We also analyzed these zircons with an infrared spectrometer. As shown in fig. 2, from zircons 9303, T9305 to 87013, both the intensity of infrared spectra and $\Delta 610$ value, 2.19, 6.57 and 20 respectively, gradually increases, indicating a decrease of the metamictization degree. As indicated in figs. 3 and 4, the peak intensity of Raman spectra is weakest in zircon 9303, and its peaks have widened considerably compared to other zircons. On the other hand, the peak intensity of Raman spectra is strongest in zircon 87013, and its peaks are distinctive and sharp. The peak intensity of Raman spectra of zircons T9305 is between these two zircons. According to the negative relationship between peak intensity of Raman spectra and metamictization degrees of zircons established by Nasdala et al. (1996), we concluded that zircons 9303 has the largest metamictization degree, but zircons 87013 has the smallest metamictization degree, and zircons T9305 have a metamictization degree between them. This conclusion is consistent with that achieved by the infrared spectrum analysis described above. In order to quantify the relationship between the metamictization degree and Raman spectra of zircons, we have measured peak intensity (H) and half-height width (W) of peak Eg ($\nu_4$) 355 cm$^{-1}$ (other peaks are also ok) in each zircon, and calculated their ratio = H/W = Δ355 as listed in table 2. Among the three groups of zircons 9303, T9305 and 87013, their average Δ355 values are 2.375, 24.67 and 26.28 respectively. Their relative size order is consistent with Δ610 achieved from their infrared spectra, which indicates that like infrared spectra, Raman spectra can also be used to quantitatively calculate the metamictization degree of zircons. Namely, the smaller the Δ355, the higher the metamictization degree, or it is the opposite. The difference between them is that the infrared spectrum data in this paper came from a mixture of multi crystals in a sample. However, the Raman spectrum data can come from a single analyzed point in a single zircon crystal.

3.3 Micro Raman spectrum zonation of zircons
3.3.1 The differences between Raman microprobe and electron microprobe analyses

Through electron microprobe (EMP) analysis, we can gain an accurate chemical composition of a crystal in a very small area of 1 x 1 μm. Therefore, we can accurately get the element zonation features of a crystal if the analyzed elements are over the detection limit of an EMP. The zircons from granites, pegmatites and some metamorphic rocks usually have a relatively high content of U and Th. In a previous study, we have found that the U + Th concentration of zircons from granites and pegmatites (magmatic origin) increase from core to rim of a crystal (Bao, 1995; Bao et al., 1996). This is consistent with the conclusions achieved by Bibikova (1979) through an investigation on the zircons from granites. However, the zircons from mafic volcanic rocks, such as the zircons T9305, M-y1 and M-y2, have a low concentration of U and Th below the detection limit of an EMP. Therefore, it is difficult to gain a reliable U and Th zonation features of these zircons. Differing from an EMP, a Raman spectrometer samples through laser ablation, not only on the surface, but also to an at least 2 μm of depth below the surface of a sample. In addition, the sampled spot's size can range from 2 μm to 10 μm. Therefore, the sampled volume of a Raman spectrometer is many times larger than that of an EMP. Consequently, its relative analysis sensitivity and accuracy in U + Th analysis is much higher than the latter. In addition, Raman spectrum analysis shows U, Th zonation features by measuring the metamictization degree in different positions in a crystal, and the metamictization degree is a reflection of crystal lattice destruction induced by not only the measurable U and Th to an EMP, but also the decayed U and Th that are existing as their daughter element - lead. An EMP can not detect the decayed U and Th, but Raman spectrum data do give us information from both the decayed and un-decayed U + Th. Therefore, Raman spectrum data can more accurately reflect the U+Th distribution pattern in a crystal right after its formation, which is favorable for accurately identifying zircon geneses.

3.3.2 Raman spectrum zonation of magmatic zircons

These Raman spectrum peaks of zircons have been assigned in a previous study (ref 5 in Bao and Gan, 1996). As shown in fig.3 and table 2, the magmatic zircons exhibit a trend of decreasing Raman spectrum peak intensities and Δ355 values from core to rim of a crystal, which indicates an increase in metamictization degree from core to rim. This corresponds to a trend of increasing U + Th concentration from core to rim of a crystal based on the direct positive relationship between metamictization degree and U + Th concentration established by Nasdala et al. (1996). This confirmed the previous conclusion that there is a general composition zonation trend of increasing U + Th concentration from core to rim of a magmatic zircon crystal (Bao, 1995).

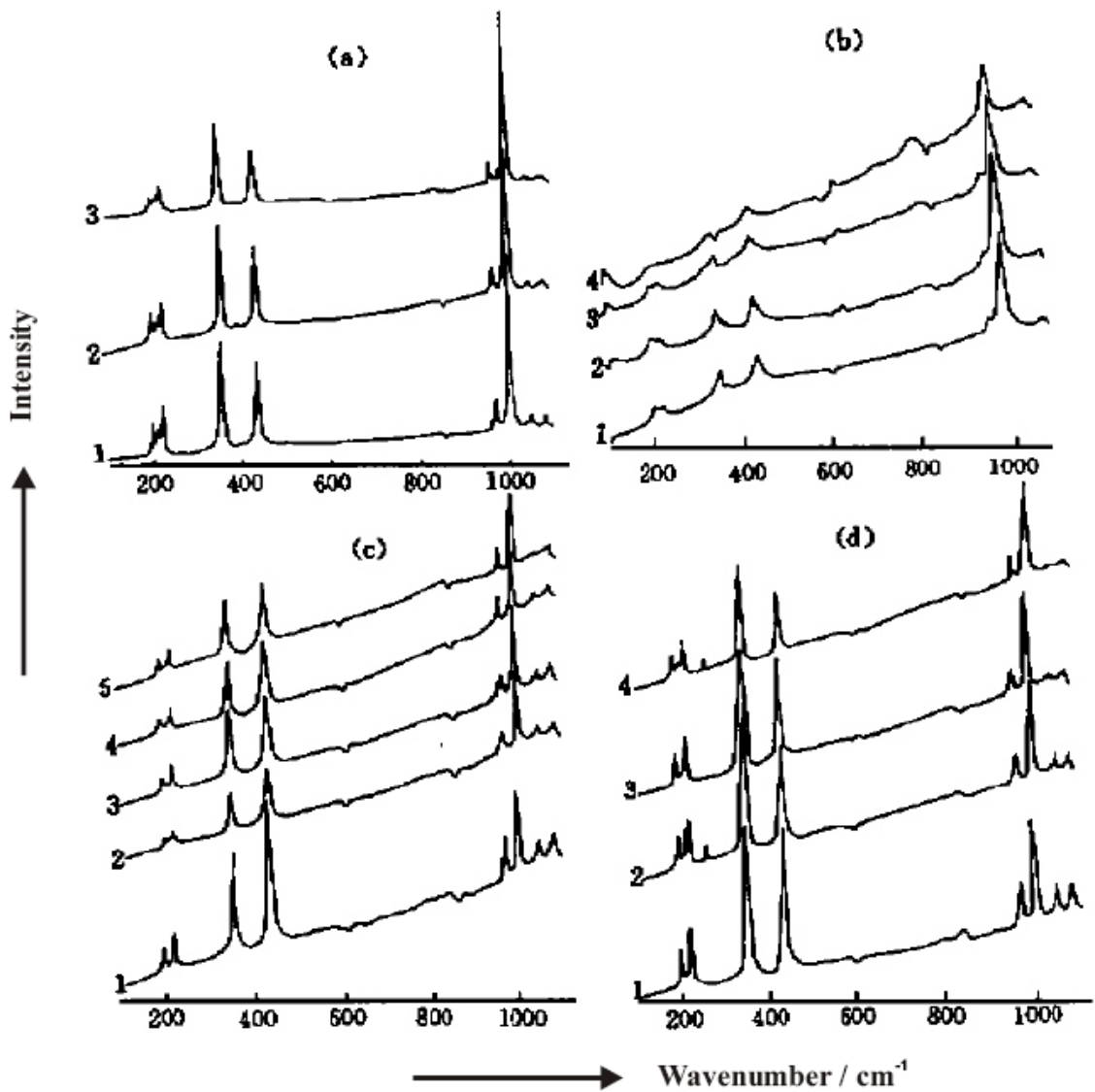

**Fig. 3　Raman spectra from core to rim (1→5) of magmatic zircons**
**(a) T9305; (b) 9303; (c) M-y1; (d) M-y2**

3.3.3 Raman spectrum zonation of metamorphic zircons

　　As shown in fig.4 and table 2, there are two kinds of metamorphic zircons according to their Raman spectrum zonation features. The first kind of metamorphic zircons, such as zircons M-b2 and 185/38, exhibits a trend of increasing both the intensities and Δ355 values of Raman spectrum peaks, which is opposite to those of the magmatic zircons as illustrated in figs. 3 and 4, and table 2. Based on the relationship between the metamictization degree and U + Th concentration, this indicates a decrease of U + Th concentration from core to rim of a crystal. This is consistent with the U + Th zonation trend of zircon 185/38 as shown in fig.1 b, and also consistent with the conclusion achieved by Bao (1995). The second kind of metamorphic zircons show no significant variation through the crystals in both the intensities and Δ355 values of Raman spectrum peaks, such as zircons 87013 and M-b1. This indicates that the U and Th concentration through a crystal has no obvious variation. From fig.4, this kind of zircons exhibit very sharp Raman spectrum peaks with large intensities. Therefore, they may not have experienced any metamictization or they may have undergone little metamictization. Zircon 87013 also exhibit very sharp infrared spectrum peaks with large intensity, as shown in fig. 2, indicating very good crystallinity. Namely, its crystal lattice has not been destroyed by the radiation of U and Th. The calculated Δ610 has a very large value of 20, and is much large than the range of < 2.0 of a metamictized zircon defined by Gao and Pang (1986). Therefore, this kind of zircon may not have experienced any or have only undergone little metamictization. Their crystals may contain no U and Th or have a trace amounts of them.

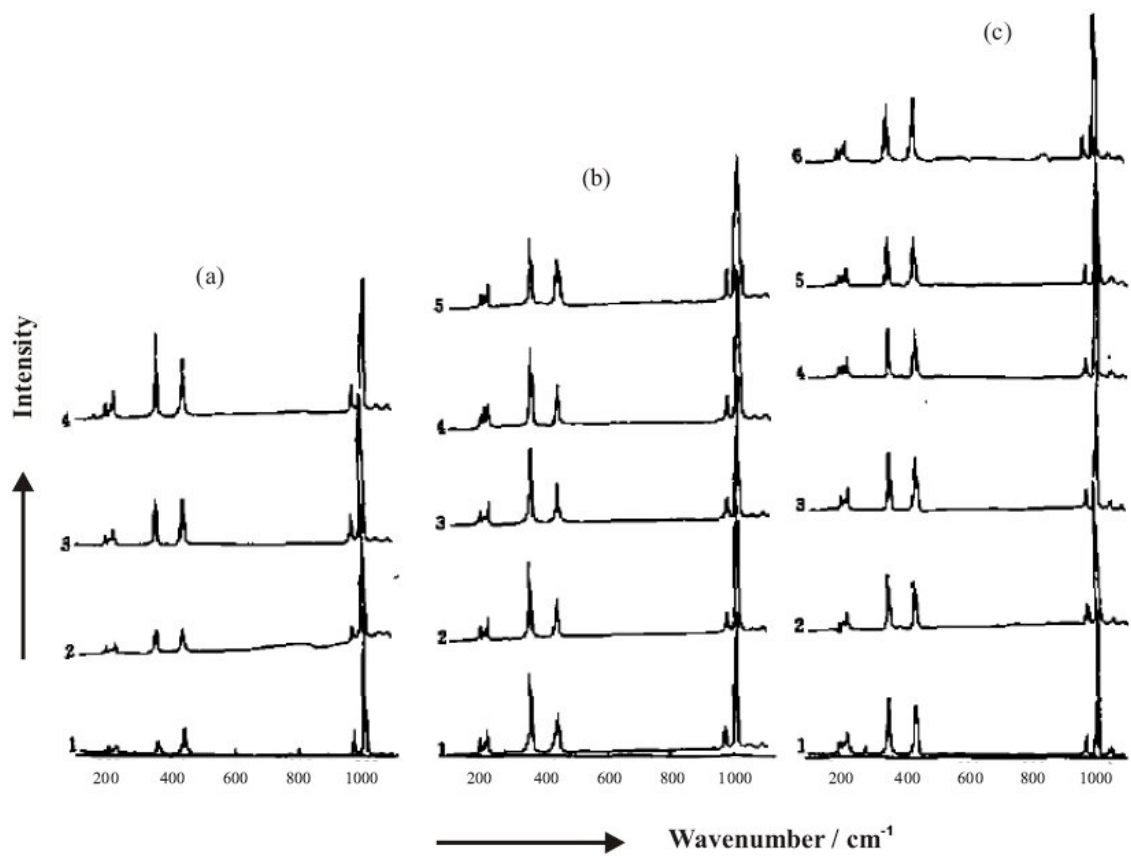

**Fig. 4  Raman spectra from core to rim (1 → 6) of metamorphic zircons**
   (a)  M-b2; (b) M-b1; (c) 87013

   **The possible formation mechanism of the low U and Th metamorphic zircons:** with an increase in metamorphic grade accompanied by an increase in temperature, the U and Th concentrations in the parent rock of zircons decrease according to Tarney et al. (1982) and Tctsumi (1988). Furthermore, the migration rate from the parent rock is

**Table 2 The intensity and half-width and their ratio values of the Raman spectra at peak (355 cm$^{-1}$) of the magmatogenic and metamorphogenic zircons***

| Genesis | Sample # | Spot # | Intensity H/mm | Half-height Width W/mm | Δ355 H/W | Genesis | Sample # | Spot # | Intensity H/mm | Half-height Width W/mm | Δ355 H/W |
|---|---|---|---|---|---|---|---|---|---|---|---|
| Magmatic | T9305 | 1 | 44.0 | 1.5 | 29.33 | Metamorphic | M-b-1 | 1 | 8.5 | 1.2 | 7.08 |
| | | 2 | 41.0 | 1.5 | 27.33 | | | 2 | 13.2 | 1.2 | 11.00 |
| | | 3 | 33.0 | 1.9 | 17.37 | | | 3 | 25.0 | 1.2 | 20.83 |
| | 9303 | 1 | 10.0 | 3.0 | 3.33 | | | 4 | 41.0 | 1.0 | 41.00 |
| | | 2 | 9.2 | 3.1 | 2.97 | | M-b-2 | 1 | 43.5 | 1.3 | 33.46 |
| | | 3 | 6.0 | 3.0 | 2.00 | | | 2 | 42.0 | 1.1 | 38.18 |
| | | 4 | 4.2 | 3.5 | 1.20 | | | 3 | 42.0 | 1.0 | 42.00 |
| | M-y-1 | 1 | 41.0 | 1.5 | 27.33 | | | 4 | 44.0 | 1.0 | 44.00 |
| | | 2 | 16.0 | 1.2 | 13.33 | | | 5 | 37.0 | 0.7 | 52.86 |
| | | 3 | 28.0 | 1.5 | 18.67 | | 87013 | 1 | 32.5 | 1.3 | 25.00 |
| | | 4 | 22.0 | 1.3 | 16.92 | | | 2 | 31.0 | 1.4 | 22.14 |
| | | 5 | 22.5 | 1.6 | 14.06 | | | 3 | 32.0 | 1.2 | 26.67 |
| | M-y-2 | 1 | 71.5 | 1.6 | 44.69 | | | 4 | 27.0 | 0.9 | 30.00 |
| | | 2 | 59.0 | 1.5 | 39.33 | | | 5 | 26.0 | 1.0 | 26.00 |
| | | 3 | 48.5 | 1.5 | 32.33 | | | 6 | 33.5 | 1.2 | 27.91 |
| | | 4 | 39.0 | 2 | 19.50 | | | | | | |

***Calculated from the original Raman spectra.**

larger, when the ionic radius is larger (Tctsumi, 1988). The ionic radius of U$^{4+}$ is 1.14 Å, Th$^{4+}$ 1.18 Å. They are much larger than 0.98 Å of Zr$^{4+}$. Consequently, in the process of metamorphic heating, U and Th will become active and gradually removed with the newly formed metamorphic fluid from the parent rock. Therefore, in the beginning stage of metamorphism, with the rise of metamorphic grade or temperature, the U and Th concentrations of the parent rock gradually decrease. The zircons crystallized in the parent rock will exhibit a zonation trend of decreasing U and Th from core to rim (Bao, 1995). At the same time, the intensity of the Raman spectrum peaks of these zircons, such as zircons 185/38 and M-b2, increases from core to rim. After approaching the top metamorphism or the late stages, most of the U and Th will have been removed from the parent rock. The zircons crystallized in these stages will have a very low or no U and Th in their crystals. This may be the reason why the intensities of their Ranman spectrum peaks show no obvious or no variation through a crystal.

Besides, it should be noted that the metamictization degree depends on not only the U and Th contents, but also the age of a zircon, since metamictization degree is a scale of crystal lattice destruction induced by the radiation from U and Th. The older the zircon crystal, the more destruction produced by this kind of radiation and the higher the metamictization degree. From this point, the U and Th concentrations estimated from the Raman spectra of a zircon can only be used within one crystal, or crystals with the same age. Different metamictization degrees among different crystals may originate from their different ages.

In a previous study, we have found that the background levels of metamorphic zircons are much lower than those of magmatic zircons. Furthermore, the background lines in the former are usually horizontal, but in the latter they are sloped (Bao and Gan, 1996). We also found this characteristic in the newly analyzed Raman spectra as shown in figs. 3 and 4. Pilz (1987) and Bao and Gan (1996) considered that a high level of background line of Raman spectra may be produced by a high concentration of fluorescent impurities. Magmatic zircons crystallized from magma with a higher temperature, therefore have larger spaces in their lattice to absorb more fluorescent impurities. A high content of fluorescent impurities produces a high background level in the Raman spectra of magmatic zircons. On the other hand, metamorphic zircons that crystallized during metamorphism with a relatively low temperature, give very small spaces to absorb fluorescent impurities. Therefore, they have a very low concentration of fluorescent impurities, which produce a low background level of Raman spectra of metamorphic zircons.

## 4. CONCLUSIONS

1) The metamictization degree through a zircon crystal can be determined by micro-Raman spectrum analysis. Therefore, the U + Th zonation feature of a zircon right after its formation can be estimated with this method. It is especially useful to those zircons crystallized in mafic magmas or metamorphic rocks with a low U and Th content.

2) Magmatic zircons exhibit a decreasing trend in both the intensity and $\Delta 355$ value of Raman spectrum peaks from core to rim of a crystal, indicating an increase trend in U + Th concentration from core to rim of a crystal right after its formation.

3) There are two kinds of crystals in metamorphic zircons according to their Raman spectra. One of these exhibits an increasing trend in both the intensity and $\Delta 355$ value of Raman spectrum peaks from core to rim of a crystal, indicating a decreasing trend in U + Th concentration from core to rim of a crystal right after its formation. Another kind of zircons exhibits none or no obvious variation in Raman spectrum intensities through a crystal, indicating that they are the metamorphic zircons with no or a very low concentration of U and Th.

Because U and Th zonation trends through a crystal of zircons can be used to identify their origins, therefore, the zonation trends through a crystal in micro Raman spectrum intensity and $\Delta = H/W$ values of zircons also can be used as a method to identify their origins.

**Notes:**

This is our paper published originally in Scientia Geologica Sinica, 1998, Vol.33 (4): 455-462.

A decreasing trend in crystallization temperature from core to rim of magmatic zircon crystals described in our papers is consistent with the calculated core-to-rim variation in crystallization temperature of a 4.3 billion year old magmatic zircon (Watson and Harrison, Science, 2005, V308, 841-844).

It is widely accepted that zircons can record the evolution history of the Earth and planets. For instance, through the studies of geochronology and isotope geochemistry of 4.0 - 4.4 billion years old zircons, it was derived that continental crusts began to form in Earth's first 100 million years (Ma) (Watson and Harrison, same as above; Harrison et al.,

Science, 2005, V310, 1947 – 1950), although they may have developed mainly around 2700 Ma and 1900 Ma ago (Condie et al., Precambrian Research, 2005, V139, 42-100).

The U, Th zonations of zircons described in our papers actually reflect the migration trends of U and Th in the Earth. U and Th are the most important heat-producing elements in the Earth and other terrestrial planets, and are the main energy sources for planetary evolution. At the same time, the X-rays released from them have an important influence on life beings. We have proposed theoretical migration models and distribution patterns of U and Th in Earth and other terrestrial planets. Their influence on the Earth and terrestrial planetary dynamics, and the origin and evolution of life (including a graphic summary of zircon composition zonations and morphologies) has also been discussed in:

Bao X., Zhang A., Geochemistry of U and Th and its influence on the origin and evolution of the Earth's crust and the biological evolution. Acta Petrologica et Mineralogica, 17(2): 160-172 (1998), or at http://arxiv.org, arXiv: 0706.1089, June 2007.

**Acknowledgements**

I would like to thank Dr. RA Secco for offering me a postdoctoral position, which provoked my interest in translating this paper in my after-work time. The original work was supported by a grant awarded by the National Natural Science Foundation of China (49202021) to Xuezhao Bao.